\documentclass[review]{elsarticle}

\usepackage{lineno,hyperref}
\modulolinenumbers[5]

\journal{Annals of Physics}

\begin{document}

\begin{frontmatter}

\title{Revisiting Vaidya-Tikekar stellar model in the linear regime}

\author[label1,correspondingauthor]{Ranjan Sharma\fnref{myfootnote}} \author[label2]{Shyam Das} \author[label3]{Megan Govender} \author[label4]{Dishant M. Pandya}

\cortext[correspondingauthor]{Corresponding author}
\fntext[myfootnote]{Email: rsharma@associates.iucaa.in}

\address[label1]{Department of Physics, Cooch Behar Panchanan Barma University, Cooch Behar 736101, India.}
\address[label2]{Department of Physics, P. D. Women's College, Jalpaiguri 735101, India.}
\address[label3]{Department of Mathematics, Faculty of Applied Sciences, Durban University of Technology, Durban 4000, South Africa.}
\address[label4]{Department of Mathematics and Computer Science, Pandit Deendayal Petroleum University, Gandhinagar 382007, Gujarat, India.}

\begin{abstract}
We obtain a new class of solutions by revisiting the Vaidya-Tikekar stellar model in the linear regime. Making use of the Vaidya and Tikekar metric ansatz [J. Astrophys. Astron. {\bf3} (1982) 325] describing the spacetime of static spherically symmetric relativistic star composed of an anisotropic matter distribution admitting a linear EOS, we solve the Einstein field equations and subsequently analyze physical viability of the solution. We probe the impact of the curvature parameter $K$ of the Vaidya-Tikekar model, which characterizes a departure from homogeneous spherical distribution, on the mass-radius relationship of the star. In the context of density-dependent MIT Bag models, we show a correlation between the curvature parameter, the bag constant and total mass and radius of some of the well-known pulsars viz., 4U 1820-30,  RX J1856-37, SAXJ 1808.4 and Her X-1.  We explore the possibility of fine-tuning these parameters based on current observational data. 
\end{abstract}

\begin{keyword}
Einstein field equations \sep Exact solution \sep Relativistic star \sep Vaidya-Tikekar ansatz.
\end{keyword}

\end{frontmatter}


\section{Introduction}
\label{sec:1}

Modelling of compact stellar objects compatible with observational data has remained one of the critical research areas in relativistic astrophysics ever since Schwarzschild\cite{Misner} obtained the first exact solution describing the exterior gravitational field of static spherically symmetric isolated object. The discovery of the pulsar in the year 1968 by Hewish {\em et al}\cite{Hewish}, which was subsequently identified as a rotating neutron star, was one of the significant milestones in the theoretical understanding of compact stars keeping in mind that a relativistic theory of neutron star was developed much earlier in the year 1939 by Oppenheimer and Volkoff\cite{OV}. 

The standard approach to study the physical behaviour of a relativistic compact star is to assume an equation of state (EOS) for the material composition of the star. Subsequently, by solving the Tolman-Oppenheimer-Volkoff(TOV) equations for the assumed equation of state(EOS), one obtains an estimate of the mass and size of the star. The technique helps to extract useful information about the gross physical properties of the star. Proper understanding of the composition and nature of particle interactions at an extremely high-density regime is the key in this approach. Alternatively, one solves the Einstein field equations for a given matter distribution and examines its physical viability. The latter approach has thus far yielded a large class of solutions out of which only a few are of physical interest. Due to the highly non-linear nature of the field equations, different simplifying techniques are invoked in such an approach. Some of the techniques that are often used to make the system of field equations tractable include the assumption of the fall-off behaviour of pressure or energy density, a reasonable choice of the mass function or the choice of a geometrically motivated form of the metric potentials. In this paper, we aim to develop and study a relativistic compact stellar model by employing the Vaidya and Tikekar (VT) metric ansatz\cite{Vaidya82} together with the choice of a linear EOS for the material composition of the star. 

The most remarkable feature of the VT spacetime is that the geometry of the $t=$constant hypersurface of the spacetime, when embedded in a $4$-Euclidean space, becomes spheroidal which is a departure from the spherical homogeneous distribution. The curvature parameter $K$, which appears in the metric potential ansatz, denotes the departure from the sphericity of $3$-space geometry. The robustness of the VT-model has been demonstrated through various investigations ranging from the search for exact solutions of the Einstein field equations, modelling of compact objects in classical gravity, dissipative collapse and higher dimensional gravity theories. Initially, for some discrete values of $K$, it was shown that the model could be used to describe a superdense star\cite{Knutsen87,Tikekar90}. Later, Maharaj and Leach\cite{leach1} successfully integrated the pressure isotropy condition resulting from the Vaidya-Tikekar ansatz to produce a family of solutions in terms of a general series solution which reduced to polynomials and algebraic functions for particular values of $K$. The work was subsequently extended by Mukherjee {\em et al}\cite{sailo1} in which they showed that the gravitational behaviour of the VT superdense star could be written in terms of Gegenbauer and hypergeometric functions. The work demonstrated for the first time that the VT model could describe neutron stars in which the energy density and pressure were related by an approximate linear equation of state (EOS). Subsequently, the VT model was utilized by many investigators to develop realistic stellar models. In particular, the VT ansatz was utilized to model the ultra-compact objects like X-ray pulsar Her X-1\cite{herx1}, millisecond pulsar SAX J 1808.4-3658\cite{sax} and many more observed pulsars. Such studies were shown to be useful to ascertain the EOS of the material composition at extreme densities. Subsequently, there have been several extensions of the original VT superdense stellar model to include anisotropic pressure and electric field, in particular (see for example \cite{Sharma07,kar1,koma1,jit1,Chatto,Paul2014}). More recently, the VT model has been generalized to higher dimensional spacetimes. It has been shown that for a given value of $K$ in $4$-D classical relativity, there exists a spectrum of analogue values $K_n$ in higher dimensions. Pure Lovelock-VT models have been developed by Khugaev {\em et al}\cite{Khug} and Molina {\em et al}\cite{Molina}. It is noteworthy that the higher dimensional isotropic VT models require that the spheroidal parameter be positive to ensure positivity of the energy density.

Motivated by numerous successful demonstrations of physical applicability of the VT model, we intend to study the model for a matter distribution which admits a linear EOS. In the modelling of a stellar object, a polytropic EOS $p = k\rho^{\gamma}$ has been extensively used in the past. The QCD inspired MIT Bag model EOS for strange stars composed of $u$, $d$ and $s$ quarks has a linear form $\rho = 3p+4B$, where $B$ is the bag constant\cite{Chodos74,Witten84,Farhi84,Alcock86}. Gondek-Rosi\'{n}ska {\em et al.} \cite{Gondek00} and Zdunik \cite{Zdunik} have independently studied compact stars obeying a linear EOS. Sharma and Maharaj\cite{Sharma07} have shown that by assuming a linear EOS, one can develop stellar models whose masses and radii are comparable with the observed pulsars. Ngubelanga {\em et al}\cite{Ngubelanga15} assumed a linear EOS to generate new solutions for a self-gravitating system . Mafa Takisa {\em et al}\cite{Mafa14} have developed a stellar model by solving the Einstein-Maxwell system for a charged anisotropic compact body where a linear EOS has been used.  

We note that the VT ansatz, when supplemented by an EOS, makes it an over-determined system and hence a new undetermined function must be incorporated into the system. This objective is accomplished by assuming the matter composition to be anisotropic, i.e., the radial pressure ($p_r$) not being equal to the tangential pressure ($p_t$). In relativistic astrophysics, there are sufficient grounds for the consideration of anisotropic stress at the high-density regime of a compact stellar interior, in particular\cite{Ruderman}-\cite{Herrera5}.
An exhaustive review of the origin and implications of local anisotropy on the gross physical properties of stellar bodies may be found in Ref.~\cite{Herrera6}. 

Our paper is organized as follows. In section \ref{sec:2}, we couch a spherically symmetric static star in the VT background spacetime and write down the Einstein field equations governing the physical variables of the system. The matching conditions necessary for the smooth connection of the interior spacetime to the vacuum Schwarzschild exterior is given in section \ref{sec:3}. In section \ref{sec:4}, we determine the bounds on our model parameters. We examine the physical viability of our solution in section \ref{sec:5}. In particular, we study the impact of spheroidal spacetime on the mass-radius relationship of the configuration. In section \ref{sec:6}, we discuss our results in the context of the MIT Bag model. A systematic investigation reveals a correlation between the curvature parameter $K$ of the VT model, the bag constant $B$ and the mass and radius of the star. We conclude by pointing out some interesting features of our model in section \ref{sec:7}.

\section{Vaidya-Tikekar model in the linear regime}
\label{sec:2}

Vaidya and Tikekar (VT) \cite{Vaidya82} demonstrated that it is possible to develop realistic compact stellar models by specifying the geometry ($G_{ab}$) of the Einstein field equations rather than specifying the matter part ($T_{ab}$) of the field equations. The interior of the VT superdense star is described by the line element
\begin{equation}
ds^2 = -e^{\nu(r)}(r)dt^2 + e^{\lambda(r)}dr^2dr^2 + r^2(d\theta^2 + \sin^2{\theta}d\phi^2),\label{metric}
\end{equation}
where
\begin{equation}
e^{\lambda(r)} = \frac{1 - K(r^2/L^2)}{1 - (r^2/L^2)}. \label{vt}
\end{equation}
The ansatz (\ref{vt}) has a clear geometric interpretation as follows. A $3$-spheroid immersed in a $4$-dimensional Euclidean flat space has the form
\begin{equation}
\frac{x^2+y^2+z^2}{L^2}+\frac{w^2}{b^2} = 1.
\end{equation}
Now, the parametrization
\begin{eqnarray*}
x = L \sin\delta \cos\theta\cos\phi,~~y = L \sin\delta \sin\theta\sin\phi,~~z = L \sin\delta \cos\theta,~~w = b \cos\delta,
\end{eqnarray*}
together with a transformation $r = L \sin\delta$ and substitution $1- b^2/L^2 = K$ leads to
\begin{equation}
d{\sigma}^2 = \frac{1-K\frac{r^2}{L^2}}{1-\frac{r^2}{L^2}}dr^2 + r^2(d{\theta}^2 + \sin^{2}\theta d{\phi}^2),\label{Eucl_eq}
\end{equation}
as the metric on the $3$-spheroid. Therefore, in Schwarzschild coordinates, the $t=$ constant hypersurface of the space-time metric
\begin{equation}
ds^{2} = - e^{\nu(r)}dt^{2}+d{\sigma}^2 = - e^{\nu(r)}dt^{2} + \frac{1-K\frac{r^2}{L^2}}{1-\frac{r^2}{L^2}}dr^2 + r^2(d{\theta}^2 + \sin^{2}\theta d{\phi}^2),
\end{equation}
will have a spheroidal geometry characterized by the parameters $L$ (which has the dimension of a length) and $K$ (which denotes departure from spherical geometry). The metric will be spherically symmetric and well behaved for $r < L$ and $K < 1$. For $K = 1$, the spheroidal $3$-space degenerates into flat $3$-space. In the case $K = 0$ (i.e., $b=L$), it becomes spherical. It is worthwhile to note that the metric with $K = 0$ and
\begin{equation}
e^{\nu(r)} = \left[A + B\left(1-\frac{r^2}{L^2}\right)^{1/2}\right]^2,
\end{equation}
leads to the Schwarzschild interior solution corresponding to an `incompressible' fluid sphere. For a spherically symmetric static configuration, we shall utilize the VT ansatz to generate new class of solutions for the metric potential $\nu(r)$ which should be well behaved and capable of describing realistic stars.

We assume that the stellar composition is anisotropic in nature and accordingly the energy-momentum tensor of the stellar fluid is taken in the form
\begin{equation}
T^a_{b} = {\mbox diag}\left(-\rho, p_r, p_t, p_t\right),\label{2}
\end{equation}
where $\rho$, $p_r$ and $p_t$ are the energy density, radial pressure and tangential pressure, respectively. The comoving fluid
four-velocity ${\bf u}$ is given by
\begin{equation}
u^a = e^{-\nu/2} \delta^{a}_0 \,.
\label{2'}
\end{equation}
The Einstein field equations for the line element (\ref{metric}) are then obtained as (in system of units having $8\pi G =1$ and $c=1$)
\begin{eqnarray}\label{g3}
\rho &=&
\frac{\left(1 - e^{-\lambda}\right)}{r^2} + \frac{\lambda^{\prime}e^{-\lambda}}{r}, \,\label{g3a} \\ \nonumber \\
p_r &=&  \frac{\nu^{\prime}e^{-\lambda}}{r} - \frac{\left(1 - e^{-\lambda}\right)}{r^2},\, \label{g3b}  \\  \nonumber \\
p_t &=& \frac{e^{-\lambda}}{4}\left(2\nu^{\prime\prime} + {\nu^{\prime}}^2  - \nu^{\prime}\lambda^{\prime} + \frac{2\nu^\prime}{r} - \frac{2\lambda^\prime}{r}\right) , \label{g3c}
\end{eqnarray}
where primes represent differentiation with respect to the radial coordinate $r$. In order to close the system of equations, we assume that the interior matter distribution obeys a linear equation of state of the form
\begin{equation}
p_r = \alpha \rho -\beta,  \label{eos}
\end{equation}
where $\alpha$ is a constant. This assumption is possible as the anisotropic nature of the fluid provides an additional degree of freedom in our construction.

Substitution of equations (\ref{g3a}) and (\ref{g3b}) into (\ref{eos}) yields
\begin{equation}
\nu = \int{r e^{\lambda}\left[\frac{(\alpha+1)(1-e^{-\lambda})}{r^2}+\frac{\alpha \lambda^{\prime}e^{-\lambda}}{r} -\beta\right]dr}.\label{gf}
\end{equation}
The problem of solving the system is now reduced to identifying a single generating function. In other words, prescribing the metric function $\lambda(r)$ as in  (\ref{vt}) gives a complete gravitational behaviour of the model. The algorithm presented by Herrera {\em et al}\cite{Herrera2008} to obtain all static spherically symmetric locally anisotropic fluid distributions is more general in the sense that it requires a single generating function together with a physically motivated ansatz (the choice of the linear EOS is just a special case).  In our case, the generating function is given by equation (\ref{vt}) and the physically motivated ansatz is the imposition of a linear EOS. These then generate $\Pi(r)=p_r-p_t$  which on using the equation  
\begin{eqnarray}
ds^2 &=& \frac{z^2(r)e^{\int{([4/r^2z(r)]+2z(r))dr}}}{r^6(-2\int{\frac{z(r)(1+\Pi(r)r^2)e^{\int{([4/r^2z(r)]+2z(r))dr}}}{r^8}}dr+C)}dr^2\nonumber\\
&&+ r^2(d\theta^2 + \sin^2{\theta}d\phi^2)-e^{\int{(2z(r)-2/r)dr}}dt^2,\label{metric1}
\end{eqnarray}
completes the model. In our case, combining equations (\ref{vt}) and (\ref{eos}), we identify the generating function $z(r)$ as
\begin{equation}
z(r) = \frac{1}{r} - \frac{n r}{L^2-r^2}-\frac{\alpha K r^2}{L^2-K r^2}-\frac{K\beta r}{2}\label{gf1}
\end{equation}   
which yields the solution in the form
\begin{equation}
e^{\nu} = A\left(1-\frac{r^2}{L^2}\right)^n \left(1-\frac{K r^2}{L^2}\right)^{\alpha}  e^{K(L^2-r^2)\beta/2},\label{soln}
\end{equation}
where $A$ is a constant of integration and 
\begin{equation}
n= \frac{1}{2}\left[-1-3\alpha+L^2\beta+K(1+\alpha-L^2\beta)\right].\nonumber
\end{equation}

Subsequently, the physical quantities are obtained as
\begin{eqnarray}
\rho &=& \frac{(1-K)(3-\frac{ K r^2}{L^2})}{L^2(1-\frac{K r^2}{L^2})^2},\label{den}\\
p_r &=& \alpha \rho-\beta,\label{radpres}\\
p_t &=& \frac{(K-1)}{4 (L^2-r^2)\left(L^2-K r^2\right)^3 \left(L^2-K R^2\right)^4}\sum_{i=1}^7 F_i,\label{tangpres}
\end{eqnarray}
where each $ F_i (i=1,2,...,7) $ is a function of parameters $K$, $\alpha$, $r$, $R$ and $L$ and are given by the following relations:
\begin{eqnarray*}
 F_1 &=& K^2 L^8 \left[25 \alpha ^2 (K-1) r^2 \left(r^2-R^2\right)^2+(K-1) r^2 \left(r^4+16 r^2 R^2+18 R^4\right)\right.\\&&
\left. -2\alpha \left((17 K+7) r^6+(57 K+35) r^4 R^2+16 K r^2 R^4-22 K R^6\right)\right],\\
F_2 &=& 2 K^3 L^6 \left[-15 \alpha ^2 (K-1) r^2 \left(r^2-R^2\right)^2 \left(r^2+R^2\right)\right.\nonumber\\
&&\left.-2 (K-1) r^2 R^2 \left(r^4+6 r^2 R^2+3 R^4\right)+\alpha  \left(3 (K+1) r^8+2 (19 K+9) r^6 R^2\right.\right. \\&&\left.\left.+(25 K+47) r^4 R^4 -8 K r^2 R^6-6 K R^8\right)\right],\\
F_3 &=& \alpha  \left[3 (K+1) r^8+2 (19 K+9) r^6 R^2+(25 K+47) r^4 R^ 4\right.\\
&&\left.-8 K r^2 R^6-6 K R^8-15 \alpha ^2 (K-1) r^2 \left(r^2-R^2\right)^2 \left(r^2+R^2\right)\right],\\
F_4 &=& K^4 L^4 r^2 \left[\alpha ^2 (K-1) \left(r^2-R^2\right)^2 \left(9 r^4+28 r^2 R^2+9 R^4\right)\right.\nonumber\\
&&\left.+(K-1) R^4 \left(6 r^4+16 r^2 R^2+3 R^4\right)\right.\\
&& \left. -2 \alpha  R^2 \left(7 (K+1) r^6+8 (3 K+2) r^4 R^2\right)+(29-9 K) r^2 R^4-4 K R^6\right],\\
F_5 &=& K^6 r^6 R^4 \left[\alpha ^2 (K-1) r^4-2 \alpha  r^2 R^2 (\alpha  (K-1)+K+1)+(\alpha +1)^2 (K-1) R^4\right],\\
F_6 &=& L^{12} \left[r^2 (-40 \alpha  K+3 K-3)+20 \alpha  K R^2\right],\\
F_7 &=& 4 K L^{10} \left[r^4 (5 \alpha + 14 \alpha  K-K+1) + r^2 R^2 (20 \alpha  K-3 K+3)-13 \alpha  K R^4\right].
\end{eqnarray*}
Note that $\beta$ is not a free parameter in this construction and can be expressed as $\beta = \alpha \rho_R$, where $R$ is the radius of the star and $\rho_R$ is the surface density given by
\begin{equation}
\rho_R=\frac{(1-K)(3-\frac{K R^2}{L^2})}{L^2(1-\frac{K R^2}{L^2})^2}.
\end{equation}
When $\rho = \rho_R$, the radial pressure vanishes i.e., $p_r(r=R) = 0$ which is an essential requirement for the development of a stellar body having finite boundary. The central density is obtained from Eq.~(\ref{den}) in the form
\begin{equation}
\rho_c = \frac{3(1-K)}{L^2},\label{cden}
\end{equation}
which shows that we must have $ K < 1 $ for positive density. Pressure is not isotropic in our construction and $S = p_t - p_r$ denotes the measure of anisotropy. The anisotropy vanishes at the centre (i.e., $S(r=0) = 0 $)  which shows the regularity of the anisotropy parameter. The mass contained within a sphere of radius $r$ is defined as
\begin{equation}
\label{massfn} m(r)= \frac{1}{2} \int\limits_0^r\omega^2
\rho(\omega)d\omega,
\end{equation}
which on integration yields
\begin{equation}
m(r) = \frac{(1-K)r^3}{2(L^2- K r^2)}.\label{mass}
\end{equation}
Clearly, the mass function is also regular at the centre i.e., $m(r=0) = 0$.

\section{Junction conditions}
\label{sec:3}
The interior solution must be matched to the Schwarzschild exterior metric 
\begin{equation}
 ds^2 = - \left(1 - \frac{2M}{r}\right) dt^2 + \left(1 - \frac{2M}{r}\right)^{-1}dr^2 + r^2 d\theta^2 + r^2 sin^2 \theta~d\phi^2,
\label{mm}
\end{equation}
across the boundary $R$. The junction conditions determine the model parameters as
\begin{equation}
L = \frac{R \sqrt{2 K M-K R+R}}{\sqrt{2M}},
\label{capital_L}
\end{equation}
where $M = m(R) $ is the total mass and
\begin{eqnarray}
 A &=& \frac{L^2-R^2}{L^2-K R^2}\left(1-\frac{R^2}{L^2}\right)^{\frac{1}{2}\left[-\alpha (K-3)+\frac{\alpha (K-1)^2 L^2 \left(K R^2-3 L^2\right)}{\left(L^2-K R^2\right)^2}-K+1\right]}\nonumber\\&&\times  \left(1-\frac{K R^2}{L^2}\right)^{-\alpha}
 \times \exp \left[-\frac{KG(K,\alpha,R, L, r)}{2 \left(L^2-K R^2\right)^2}\right], \label{x}
\end{eqnarray}
where
\begin{eqnarray}
G(K,\alpha,R, L, r) = 2 \left(L^2-K R^2\right)^2 \left(L^2-r^2\right)^{\left[\frac{\alpha  (K-1) \left(K R^2-3 L^2\right)}{2 \left(L^2-K R^2\right)^2}\right]}\nonumber\\
 +\alpha  (K-1) r^2 \left(K R^2-3 L^2\right)-\alpha  (K-1) R^2 \left(K R^2-3 L^2\right). 
\end{eqnarray}

\section{Bounds on the model parameters}
\label{sec:4}
\begin{figure}
    \includegraphics[width=0.9\columnwidth]{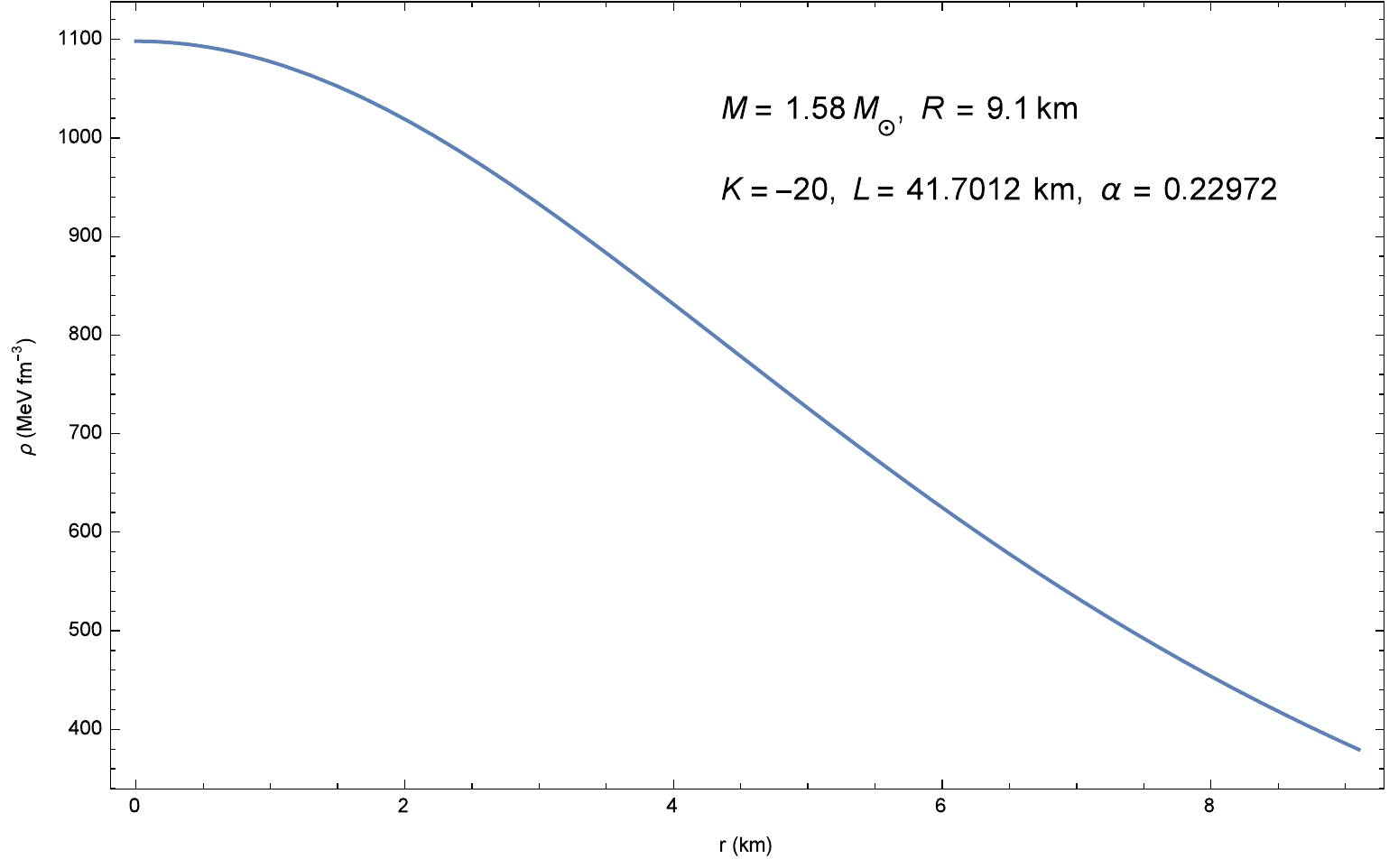}
    \caption{Fall-off behaviour of energy density.}
    \label{fg1}
\end{figure}

\begin{figure}
    \includegraphics[width=0.9\columnwidth]{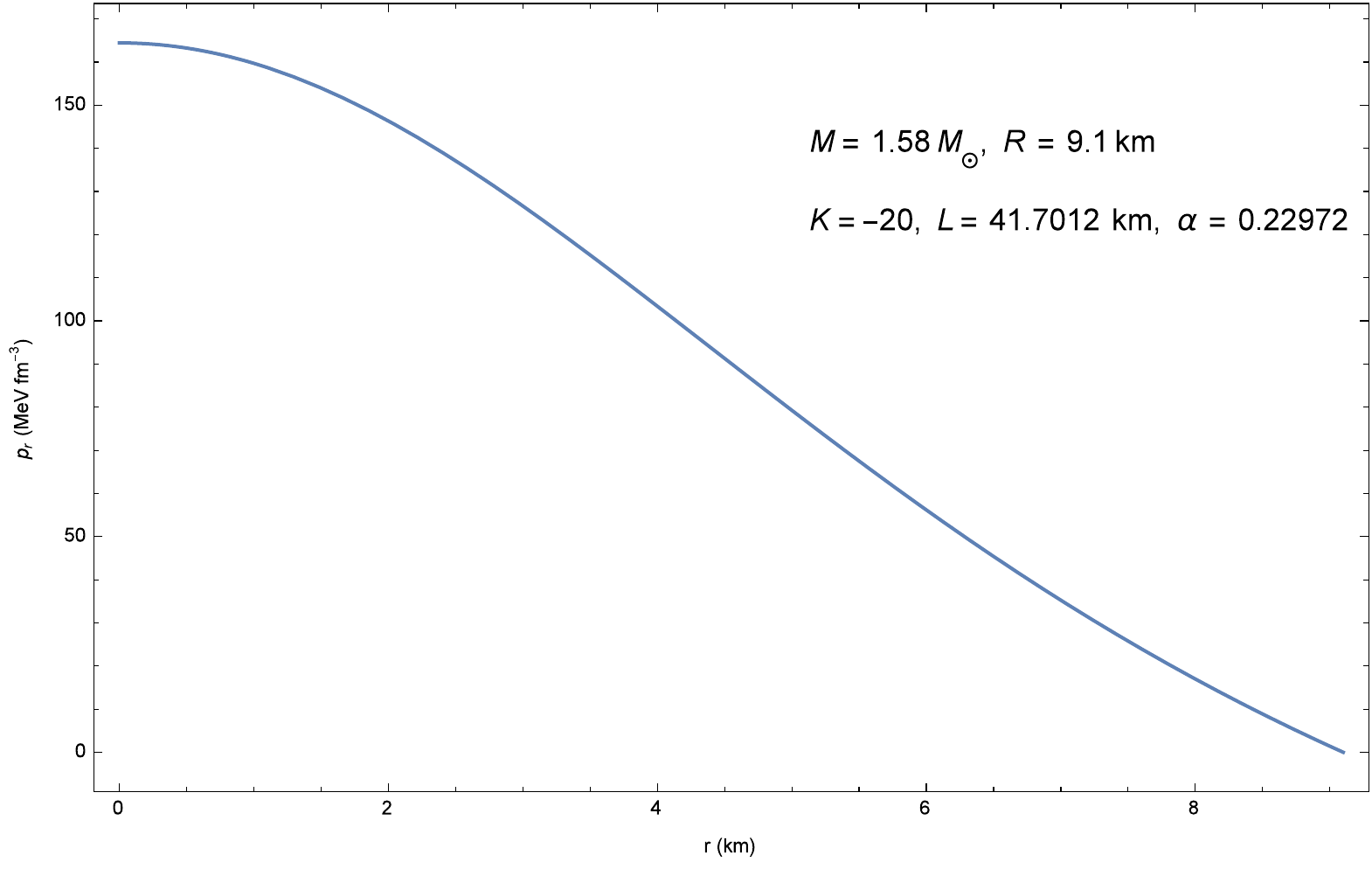}
    \caption{Fall-off behaviour of radial pressure.}
    \label{fg2}
\end{figure}

\begin{figure}
    \includegraphics[width=0.9\columnwidth]{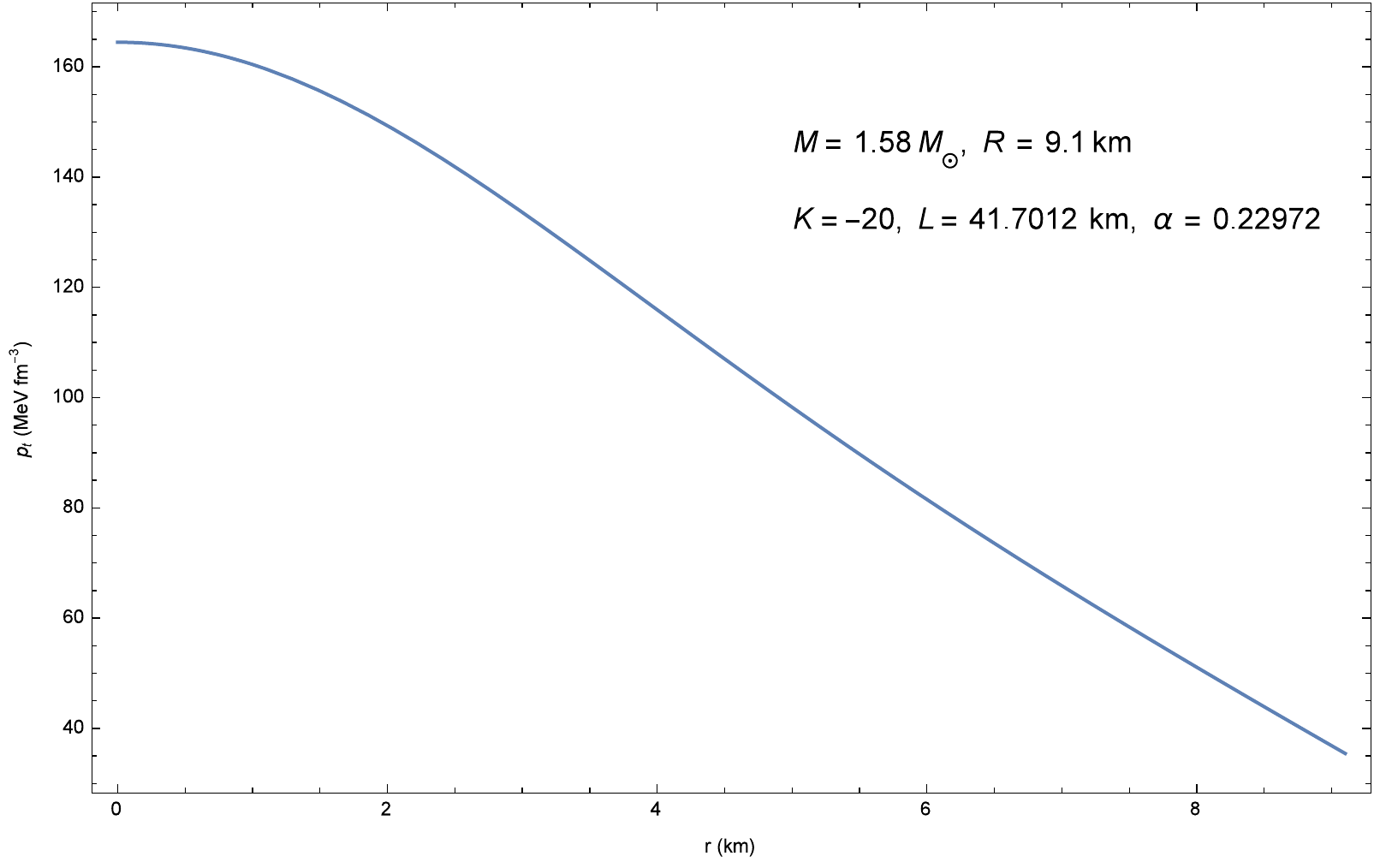}
    \caption{Fall-off behaviour of tangential pressure.}
    \label{fg3}
\end{figure}

For a physically acceptable stellar model the following conditions should be satisfied \cite{Delgaty}:
(i) $\rho > 0$, $p_r > 0$, $p_t > 0$; (ii) $\rho' < 0$, $p_r' < 0$, $p'_t < 0$; (iii) $ 0 \leq \frac{dp_r}{d\rho} \leq 1$ and (iv) $\rho-p_r -2p_t > 0$. In addition, it is expected that the solution should be regular and well-behaved at all interior points of the stellar configuration. All these requirements provide some effective bounds on the model parameter as followss:
\begin{enumerate}
\item \textbf{Regularity conditions:}
 \begin{enumerate}
 \item $ \rho (r) \geq 0, ~~ p_r (r) \geq 0, ~~ p_t (r) \geq 0 $ for $ 0
\leq r \leq R $.\\
From Eq.~(\ref{den}), we note that density remains positive if $ K < 1 $. Eq.~(\ref{radpres}) shows that for non-negative pressure we must have $ \alpha > 0 $ as well as $ K < 0 $. From equation (\ref{tangpres}), we have
\begin{equation}
p_t(r = 0) = \frac{\alpha (1-K) K R^2 \left(3 K R^2-5 L^2\right)}{\left(L^3-K L R^2\right)^2}.
\label{tangpresatcentre}
\end{equation}
We note that for $ L > R,~0 < \alpha \leq 1 $ and $ K < 0 $, the above requirement is satisfied at the centre $r=0$. For a specific set of model parameters, fulfillment of the above requirements throughout the star has been shown by graphical representation in Fig.~(\ref{fg1})-(\ref{fg3}).
\item $ p_r (r = R) = 0. $ \\
From Eq.~(\ref{radpres}), we note that the radial pressure vanishes at the boundary $R$ if we set $\beta = \alpha \rho_R $, where $\rho_R $ is the surface density.
\end{enumerate}

\item \textbf{Causality condition:}
The causality condition demands that $ 0 \leq \frac{dp_r}{d\rho} \leq 1$ at all interior points of the star.\\
Since $\frac{dp_r}{d\rho} = \alpha $, we must have $ 0 \leq \alpha \leq 1 $.

\item \textbf{Energy condition:}
For an anisotropic matter distribution, the strong energy condition $ \rho - p_r - 2p_t \geq 0 $ has to be satisfied within the stellar interior. We have at  $r = 0$
\begin{eqnarray}
\rho - p_r - 2 p_t = \frac{3(K-1)}{L^2}\frac{\left((3 \alpha -1) K^2 \frac{R^4}{L^4}+(2-5 \alpha ) K \frac{R^2}{L^2}-1\right)}{\left(1-K \frac{R^2}{L^2}\right)^2},
\label{strngenergcondatctr}
\end{eqnarray}
and at $r = R$
\begin{eqnarray}
\rho - p_r - 2 p_t&=& \frac{(1-K)}{{2 L^8}}\frac{1}{(1-\frac{R^2}{L^2})(1 - K \frac{R^2}{L^2})^3} \times \nonumber\\
&&\left[K^2 R^6 (-4 \alpha +K-3)-L^4 R^2 (5 (4 \alpha +1) K+9) \right. \nonumber\\
&&\left. +2 K L^2 R^4 (10 \alpha +(2 \alpha -1) K+6)+6 L^6\right].
\label{strngenergcondatbdry}
\end{eqnarray}
It turns out that the energy condition is satisfied if we have the bound $\frac{5 \alpha  L^2-2 L^2}{2 (3 \alpha -1) R^2}-\frac{1}{2} \sqrt{\frac{25 \alpha ^2 L^4-8 \alpha  L^4}{(3 \alpha -1)^2 R^4}}\leq K < 0 $.

\item \textbf{Monotonic decrease of density and pressure:}
A realistic stellar model should have the following properties:
 $ \frac{d\rho}{dr} \leq 0,~\frac{dp_r}{dr} \leq 0,$ for $ 0 \leq r \leq R $.\\
Now, $\frac{dp_r}{dr} = \alpha (\frac{d\rho}{dr}) = \alpha \frac{2 K r (1-K)\left(5-K \frac{r^2}{L^2}\right)}{L^4\left(1-K \frac{r^2}{L^2} \right)^3}\leq 0 $ since $ K < 0 $. It shows that both density and radial pressure decrease radially outward.
\end{enumerate}

\section{Physical viability}
\label{sec:5}
\begin{figure}
    \includegraphics[width=0.9\columnwidth]{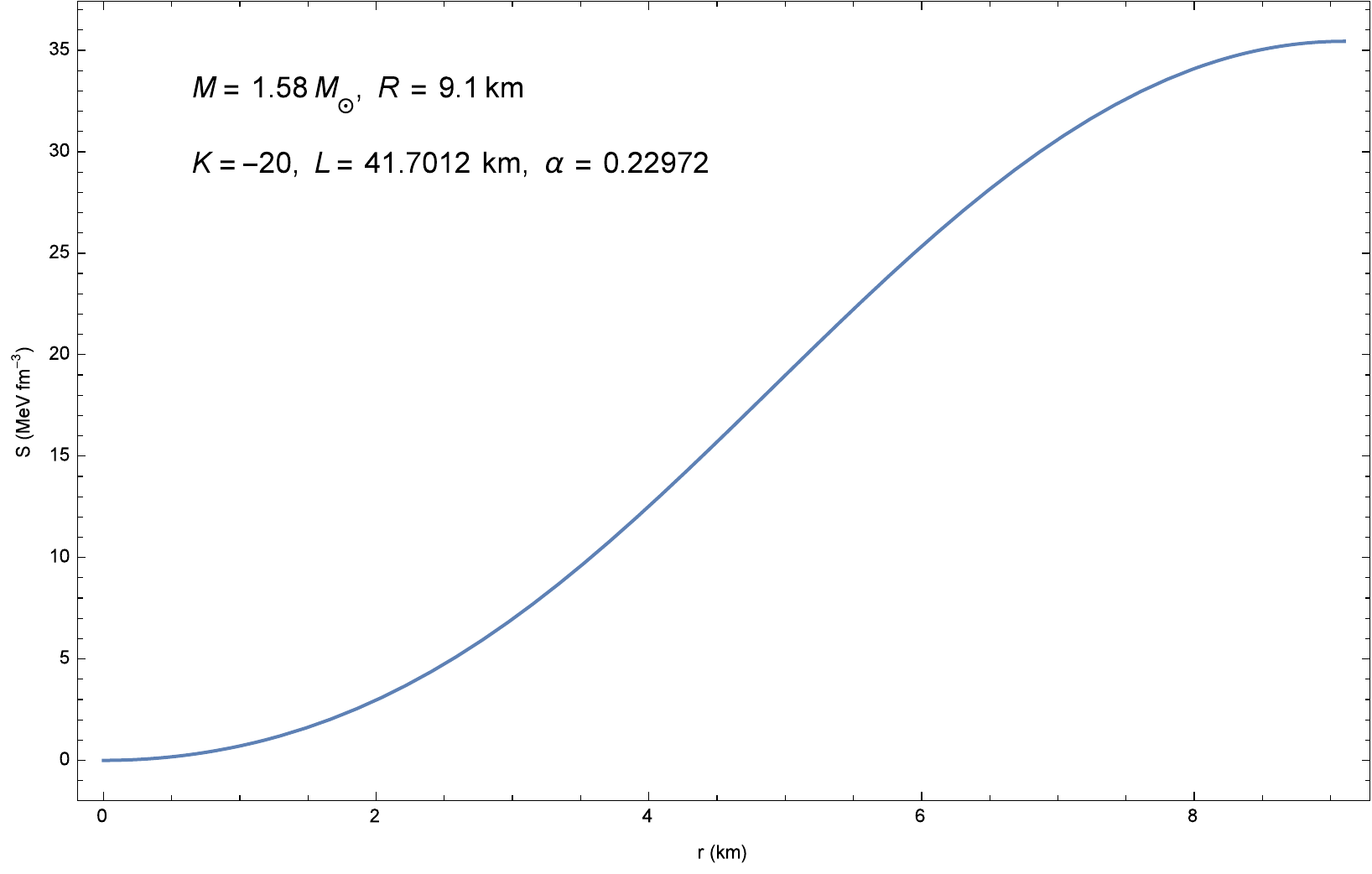}
    \caption{Radial variation of anisotropy.}
    \label{fg4}
\end{figure}

To illustrate that the solution can be used as a viable model for observed astrophysical sources, we consider the pulsar $4U 1820-30$ whose mass and radius are estimated to be $M = 1.58~M_\odot $ and $R = 9.1~$km, respectively\cite{Gangopadhyay13}. For the estimated mass and radius, the values of the constants are fixed for some assumed values of $K$ and $\alpha$ which remain as free parameters in this formulation (we have assumed $K = -20$ and $\alpha = 0.22972 $). The evaluated values of the model parameters are to study the behaviour of the physical quantities within the stellar interior. Fig.~(\ref{fg1}) - (\ref{fg3}) show the radial variation of density $\rho$, radial pressure  $p_r$ and tangential pressure $ p_t $, respectively. All the quantities decrease monotonically from the centre towards the boundary. Variation of anisotropy is shown in fig.~\ref{fg4}.

\begin{figure}
    \includegraphics[width=0.9\columnwidth]{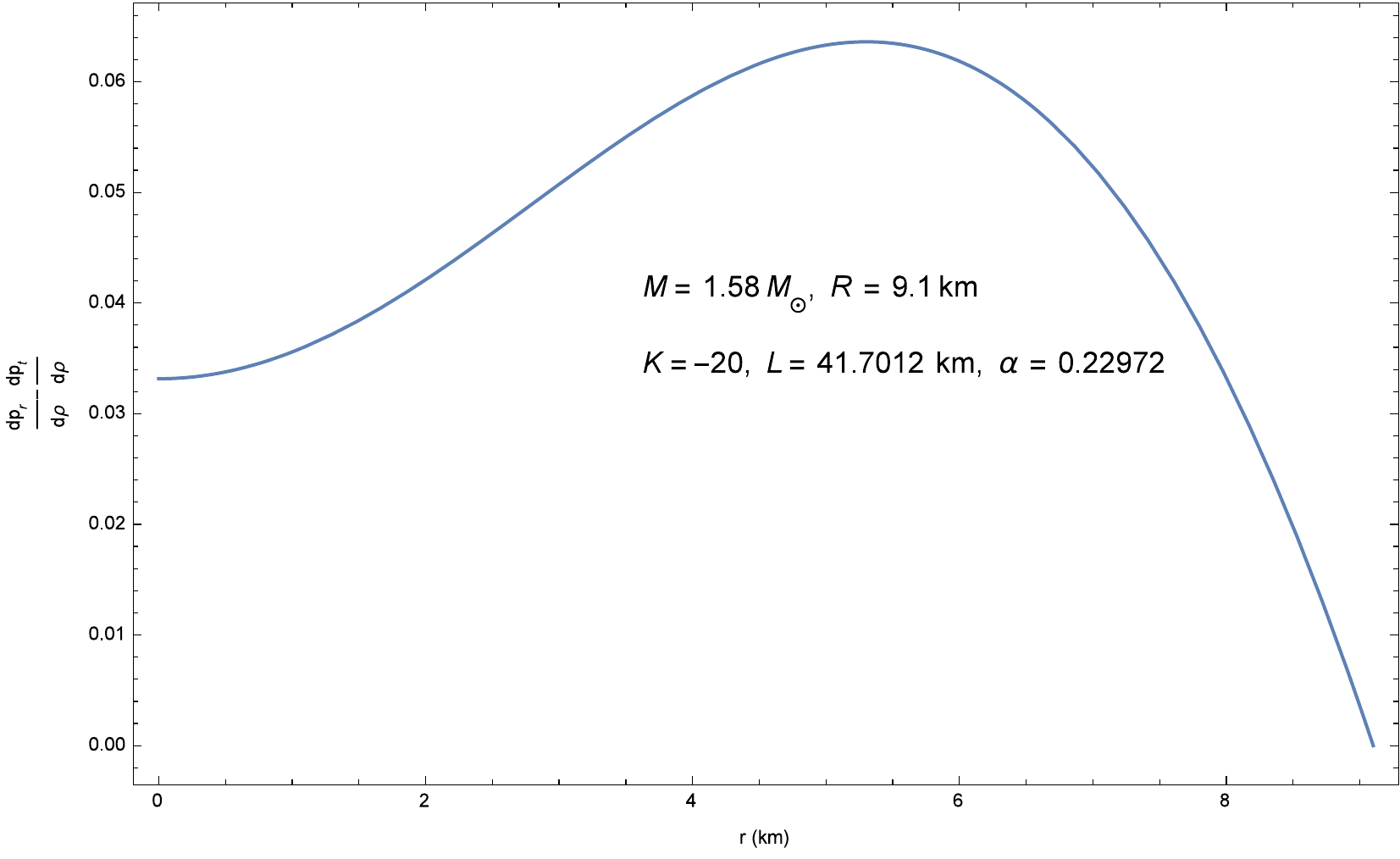}
    \caption{Fulfillment of stability requirement.}
    \label{fg5}
\end{figure}

\begin{figure}
    \includegraphics[width=0.9\columnwidth]{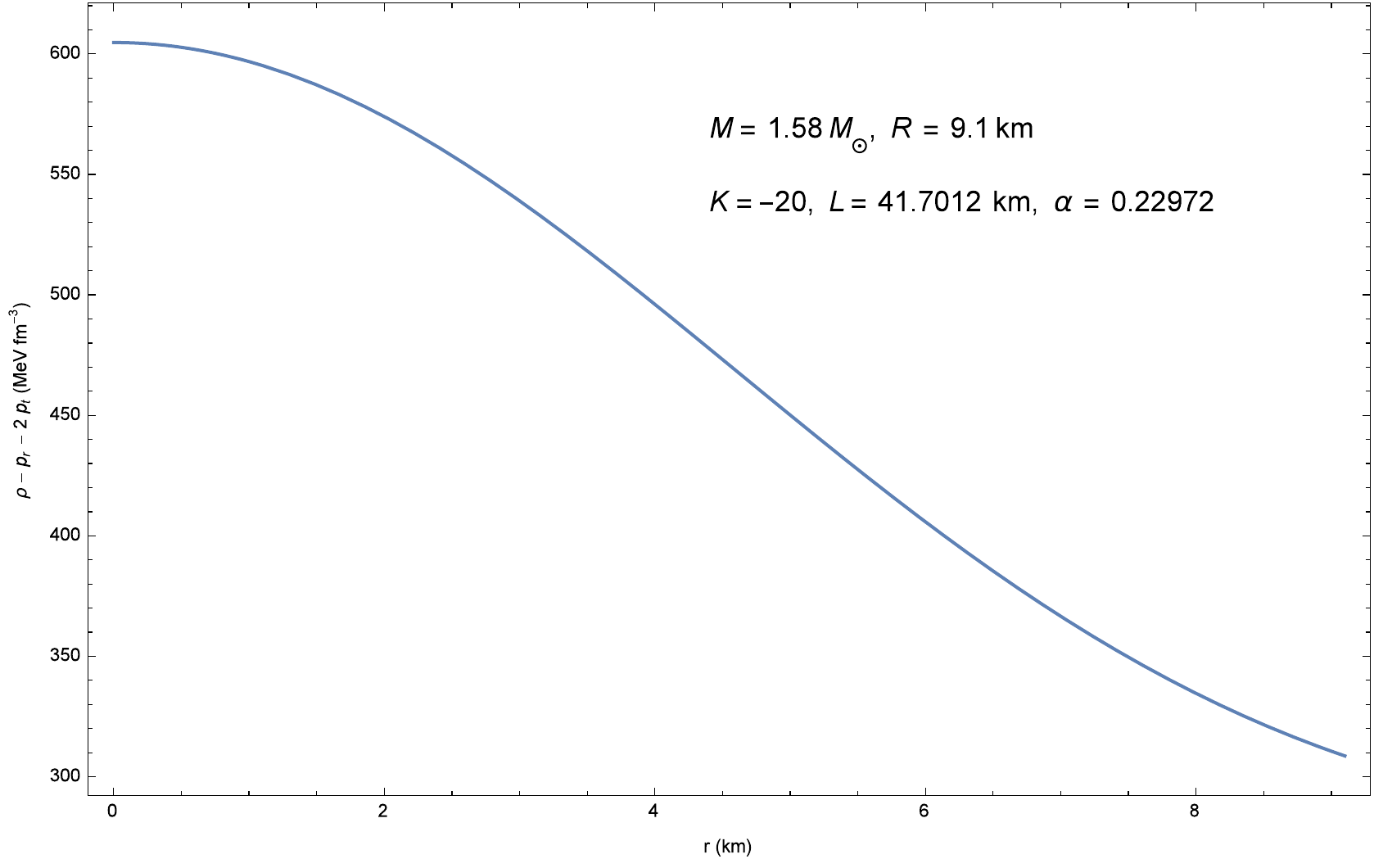}
    \caption{Fulfillment of strong energy condition.}
    \label{fg6}
\end{figure}

To examine the stability, we follow the technique used by Herrera {\em et al}~\cite{Herrera92} which states that for a stable configuration we must have $ 0 \leq \left(\frac{d p_r}{d\rho} - \frac{d p_t}{d\rho}\right) \leq 1 $. Figure \ref{fg5} clearly indicates that it is possible to find a set of values for which the configuration remains stable. The strong energy condition is shown to be satisfied in Fig.~\ref{fg6}.

\subsection{Mass-radius relationship}
\begin{figure}
    \includegraphics[width=0.8\columnwidth]{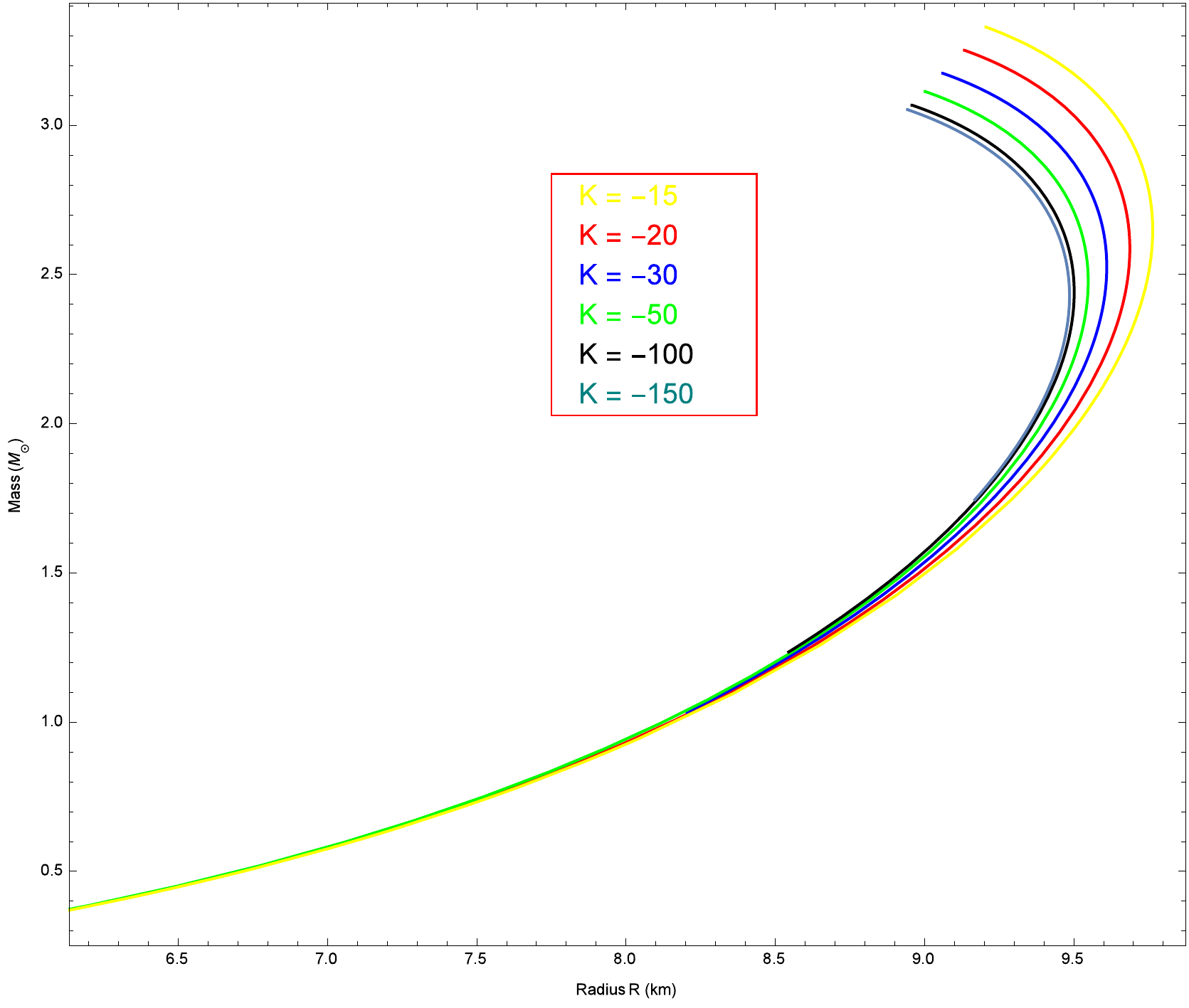}
    \caption{Mass-radius ($M-R$) relationship for different $K$ values. Assumed surface density $\rho_R = 6.77\times10^{14}~$gm cm$^{-3}$.}
    \label{fg7}
\end{figure}

We now examine the impact of departure from spherical geometry on the mass-radius relationship of a compact star. We obtain the mass-radius relationship for different values of the spheroidal parameter $K$ for a fixed surface density (see fig.~\ref{fg7}). The plot indicates that within a given radius the total mass decreases as the value of $|K|$ increases. In other words, the stellar compactness decreases as we move from sphericity of $3$-space geometry. 

\section{Bag model analogy}
\label{sec:6}
As far as compact stars are concerned, many exotic phases of matter may exist at the interior of such stars\cite{Weber2005}. In particular, the conjecture that quark matter might be the true ground state of hadrons has led to the discussion of an entirely new class of compact stars known as strange stars composed of $u$, $d$ and $s$ quarks\cite{Witten84,Farhi84,Alcock86}. In 1974, Chodos {\em et al}\cite{Chodos74} proposed a phenomenological model for quark confinement, known as the MIT bag model, in which quarks were assumed to be confined in a bag by the universal pressure `B', called the bag pressure on the surface of the bag. The value of the bag constant can be interpreted in terms of the energy difference between free-quarks and interacting quarks. In the early Bag models, two sets of values of the bag constant, namely $55$~MeV~fm$^{-3}$ and $90$~MeV~fm$^{-3}$ had been proposed. Subsequently, it was shown that a stable quark configuration was possible for $B \approx 58~$MeV~fm$^{-3}$\cite{Buballa96,Buballa99}. Later on, many QCD inspired alternative descriptions of quark confinement mechanism have been developed. As a consequence of these developments, one finds that a wider range of stability window for `B' is possible if the bag constant is assumed to be a dependent function of density/temperature/magnetic field\cite{Adami,Blaschke,Burgio,Isayev,Yaz,Bordbar,Paulucci,Avellar}. In the context of our solution, we note that the Bag model EOS 
\begin{equation}
p_r = \frac{1}{3}(\rho - 4B),
\end{equation}
can be regained simply by setting $\alpha=1/3$ and $\beta=4\alpha B$.    The model then allows us to fix the value of the bag constant or other stellar observables. To demonstrate this, we consider the isolated pulsar RX J1856-37 which has been claimed to be a strange star having mass and radius $M=0.9\pm 0.2~M_{\odot}$ and $b=6^{-1}_{+2}~$km, respectively\cite{Pons}. It should be stressed that, in general, greater uncertainty is involved in the estimation of the radius of a pulsar as compared to its mass measurement. Accordingly, for the estimated mass of RX J1856-37 in Table~\ref{tab:1}, we show the dependence of the bag constant $B$ as well as the curvature parameter $K$ on the radius of the star. In the table, the set I shows that for the standard value of the bag constant ($B=58$~Mev~fm$^{-3}$), the radius of the star is $\sim 9.7 - 9.5~$km for different choices of the curvature parameter $K$. If, however, we consider a model-dependent bag constant (we have assumed $125$~Mev~fm$^{-3}$), the radius of the star reduces to $\sim 7.4~$km for an assumed curvature parameter $K = -2$. It decreases marginally when the value of the curvature parameter is increased and takes the value $\sim 7.14~$km for higher values of the curvature parameter (e.g., $K=-100$) as can be seen in set II. Most interestingly, to have a radius $R=6~$km for the given mass, the bag constant should be $\sim 198-225~$~Mev~fm$^{-3}$ as shown in set III. Whether such high values of the bag constant are admissible from the stability point of view, however, remains beyond the scope of this investigation.    

We perform similar calculations for some other pulsars whose masses have been claimed to be well constrained in the recent past. We consider the X-ray pulsar Her X-1 and milli-second pulsar SAX J1808.4-3658  whose estimated masses are $0.85\pm 0.15~M_{\odot}$ \cite{Abu} and $0.9\pm 0.3~M_{\odot}$ \cite{Ele}, respectively. Assuming these pulsars to be governed by the MIT Bag model EOS, we examine the impacts of the curvature parameter $K$ and bag constant $B$ on their respective radii which have been compiled in Table~\ref{tab:2}. The results reveal that for the standard choice of the bag constant ($B=58$~Mev~fm$^{-3}$), one obtains radii which are on the higher side as compared to the estimations given in Ref.~\cite{Gangopadhyay13}. It turns out that to fit the observed masses and radii of these pulsars, the value of the bag constant should be much higher. If, however, one sticks to the standard value of the bag constant, the compactness of these pulsars seems to differ from the predicted values one obtains by employing other techniques.

\begin{table*}
\caption{Estimation of radius of the pulsar RX J1856-37 for different values of $K$ and $B$.}
\label{tab:1}      
\begin{tabular}{|l|c|c|c|c|c|c|r|}
\hline
Object & Mass & $\alpha$  & $B$ & $\beta$ & $K$ & $L$ & $R$ \\ 
       & ($M_{\odot}$) &   & (MeV~fm$^{-3}$) & (MeV~fm$^{-3}$) & & (km) & (km) \\ \hline 
RX J1856-37 & 0.9 & $\frac{1}{3}$ & 58 & 77.33 & -2 & 29.06 & 9.71  \\
(Set I) &  &  &  & & -10 & 51.72 & 9.53 \\ 
& & & & & -20 & 108.82 & 9.47 \\ 
& &  & & & -100 & 152.64 & 9.47 \\ \hline
RX J1856-37  & 0.9 &  $\frac{1}{3}$ & 125 & 166.67 & -2 & 18.68 & 7.40\\
(Set II) &  & & & & -10 & 33.16 & 7.21\\ 
 &  &  & & & -20 & 43.50 & 7.18 \\ 
& & & & & -100 & 93.60 & 7.14 \\ \hline
RX J1856-37 & 0.9 &  $\frac{1}{3}$ & 224.47 & 299.30 & -2 & 13.12 & 6.0\\
(Set III) & & & 200.92 & 267.90 & -20 & 31.44 & 6.0\\
& & & 197.81 & 263.75 & -100 & 67.94 & 6.0\\ \hline
\end{tabular}
\end{table*}

\begin{table*}
\caption{Estimation of radii of the pulsars SAXJ 1808.4 and Her X-1 for different values of $K$ and $B$.}
\label{tab:2}      
\begin{tabular}{|l|c|c|c|c|c|c|r|}
\hline
Object & Mass & $\alpha$  & $B$ & $\beta$ & $K$ & $L$ & $R$ \\  
 & ($M_{\odot}$) &   & (MeV~fm$^{-3}$) & (MeV~fm$^{-3}$) & & (km) & (km) \\ \hline 
SAXJ 1808.4 & 0.9 & $\frac{1}{3}$ & 58 & 77.33 & -2 & 34.13 & 10.73  \\
(Set IV) &  &  &  & & -200 & 255.74 & 10.49 \\ \hline
SAXJ 1808.4  & 0.9 &  $\frac{1}{3}$ & 100.49 & 133.99 & -2 & 21.22 & 8.0 \\
(Set V) &  & & 91.92 & 122.57 & -200 & 161.12 & 8.0\\ \hline
Her X-1 & 0.85 &  $\frac{1}{3}$ & 58 & 77.33 & -2 & 29.26 & 9.53 \\
(Set VI) & & &  &  & -200 & 217.65 & 9.31 \\ \hline
Her X-1 & 0.85 & $\frac{1}{3}$ & 92.50 & 123.342 & -2 & 22.46 & 8.1\\ 
(Set VII) & & & 85.203 & 113.65 & -200 & 171.69 & 8.1 \\ \hline
\end{tabular}
\end{table*}

\section{Discussion}
\label{sec:7}
In this work, making use of the VT metric ansatz, we have provided a new class of interior solutions describing a static and spherically symmetric anisotropic matter distribution which admits a linear EOS. The solution has been shown to be regular and well-behaved throughout the stellar configuration. The solution has been used to study the impact of deviation from sphericity of $3$-surface geometry on the mass-radius relationship vis-a-vis compactness of a star. For a strange star, the model has been utilized to constrain the bag constant. The values of the bag constant lie within the predicted range of density/temperature/magnetic field-dependent bag models. To conclude, the model developed in this paper may be used to analyze the impact of geometry on the gross physical properties of a relativistic compact star and also to fine-tune some of the observables like mass and radius of a star.

\section{Acknowledgements}
The work of RS is supported by the MRP grant F.PSW-195/15-16 (ERO) of the UGC, Govt. of India. RS gratefully acknowledges support from the Inter-University Centre for Astronomy and Astrophysics (IUCAA), Pune, Govt. of India, under its Visiting Research Associateship Programme.

\pagebreak


\end{document}